\documentclass[3p,12pt]{elsarticle}
\usepackage{lineno,hyperref,booktabs,amsmath,float,amsfonts,graphicx}
\modulolinenumbers[5]

\journal{Physica A}

\begin{document}

\begin{frontmatter}

\title{Recurrence measures and transitions in stock market dynamics}
\tnotetext[mytitlenote]{Fully documented templates are available in the elsarticle package on}

\author[mymainaddress]{Krishnadas M.}

\author[mymainaddress]{K. P. Harikrishnan}


\author[mysecondaryaddress]{G. Ambika\corref{mycorrespondingauthor}}
\cortext[mycorrespondingauthor]{Corresponding author}
\ead{g.ambika@iisertirupati.ac.in}

\address[mymainaddress]{The Cochin College, Kochi-682002, India}
\address[mysecondaryaddress]{Indian Institute of Science Education and Research (IISER) Tirupati, Tirupati-517507, India}

\begin{abstract}
The financial markets are understood as complex dynamical systems whose dynamics is analysed mostly using nonstationary and brief data sets that usually come from stock markets. For such data sets, a reliable method of analysis is based on recurrence plots and recurrence networks, constructed from the data sets over the period of study. In this study, we do a comprehensive analysis of the complexity of the underlying dynamics of 26 markets around the globe using recurrence based measures. We also examine trends in the nature of transitions as revealed from these measures by the sliding window analysis along the time series during  the global financial crisis of 2008 and compare that with changes during the most recent pandemic related lock down. We show that the measures derived from recurrence patterns can be used to capture the nature of transitions in stock market dynamics. Our study reveals that the changes around 2008 indicate stochasticity driven transition, which is different from the transition during the pandemic.
\end{abstract}


\end{frontmatter}


\section{Introduction}
The dynamics of the global financial system attracted a lot of research activities in recent decades. Following the stock market crisis of October 19, 1987, there has been a surge in interest in using approaches based on nonlinear dynamics and time series analysis in the financial markets. The global currency crisis of 1998, the dot-com bubble burst in 2001, the global financial crisis (GFC) of 2008, the European debt crisis of 2012, the Chinese crisis of 2015-2016, and the recent Covid 19 related changes after 2020 are just a few examples of events of concern from the past 20 years. It is understood that a strong abrupt volatility in stock prices causes dramatic trend shifts in a number of equities, and they continue to have a significant impact on the global economy, generating instability when faced with  normal and natural disasters. Such issues often demand a paradigm crisis in modelling, forecasting, and interpretation of socioeconomic reality. The reported research in recent years testifies that the same procedures and criteria employed in the study of natural phenomena can be applied to the study of such occurrences in socio-economic systems.  
\\Despite the fact that complexity of financial time series typically can arise from deterministic and stochastic origins, it is common practice to model them using mathematically computable models \cite{platen2001minimal}. The majority of macroeconomic contributions are based on DSGE (dynamic stochastic general equilibrium) models. These models try to explain the economy by assuming that the set of prices is the consequence of a perfect rational equilibrium. However, during the crisis events, these macro-financial models reveal some severe flaws, and they were unable to account for the observed big instability in financial markets and the macro-economy. Hence recently the focus has shifted to deterministic nonlinear models and tools and dynamical systems approach in this area also. In particular, recurrence plots and recurrence quantification analyses are gaining significance in research related to financial markets.  \cite{RePEc:eee:jeborg:v:54:y:2004:i:4:p:483-494,holyst2001observations,kyrtsou2005complex}. The techniques derived from them, are effective and powerful since they do not require any prior assumptions about the statistical properties of the time series and is resilient to non-stationarity. Moreover they capture the complexity and its variations in the dynamics underlying the system. \\

The recurrence based methods were developed over the past two decades as a new way of describing complexity of dynamics \cite{marwan2007}. The study of recurrence patterns in the data most often starts with reconstruction of the state space trajectory using delay coordinates \cite{Ambika2020}. This is a widely used technique that allows us to estimate dynamical invariants by constructing a topologically equivalent dynamical trajectory of the original dynamics from the measured (scalar) time series or data. From the dynamical trajectory, information on the recurrences of states in the dynamics is derived, capturing the important characteristics of the underlying dynamics as a 2-dimensional image, called recurrence plot(RP). The quantification of patterns in the RP using various complexity measures provides further insights into the system's dynamics. The measures from RPs are used to investigate a variety of markets, including currency exchange rates \cite{belaire2002assessing} and electricity pricing \cite{strozzi2007application}.
The characteristics of the dynamics can be studied in a complementary approach using the framework of complex networks. This has been effectively used to describe the causal signatures in seismic activity \cite{davidsen2006earthquake}, classification of binary stars \cite{george2019classification}, palaeo-climate data \cite{donges2011identification} and ECG data \cite{kachhara2019bimodality}. In the present study, we employ the framework of recurrence plots and networks, which are built from the recurrence patterns of the dynamics on the reconstructed state space trajectory to study the underlying dynamics of stock markets.\\

Many complex dynamical systems, such as climate and ecological systems, are found to undergo critical transitions or tipping \cite{scheffer2001catastrophic}. In the context of financial markets, financial meltdowns can be considered as similar transitions since they involve abrupt state changes and a slow return to the former state. In recent studies on early warning signals for tipping, financial crises are mentioned as one of the potential applications \cite{May2008}. However, early warning signals (EWS) of systemic vulnerabilities in financial markets have mostly been developed using statistical models \cite{oh2006early,reinhart2000assessing,edison2003indicators}. Scheffer et al.\cite{scheffer2009early} and Battiston et al.\cite{battiston2016complexity} suggested that financial crises are examples of critical transitions, and may indicate  "critical slowing down" before the transitions. But recent studies report that, some of the major markets are not showing any signs of critical slowing down \cite{guttal2016lack}. In this study we employ recurrence based measures from data to understand the nature of traditions in stock markets during  the financial crisis of 2008 and the most recent pandemic related lock down.\\

We start with the hypothesis that the data of each stock market is generated by a complex and nonlinear dynamical system. We try to capture the  underlying dynamics as well as variations or transitions in the dynamics based on information about the recurrences in the reconstructed dynamical trajectory. The measures from recurrence plots and networks together are used to understand the dynamics of the markets. To analyse changes in dynamics, we perform a sliding window analysis of the time series and determine recurrence measures which could aid in real-time stock market monitoring. Our results are supported by the trends in generic statistical measures like variance. The primary goal of this sliding window analysis is to comprehend the changes that occurred in the aftermath of the 2008 financial crisis and look for possible early warning signals that could detect stock market crashes or bubbles. \\

We find, at the vicinity of a critical transition, majority of the stock markets exhibits similar patterns, while some of them show delayed changes. We extend the analysis to the changes in dynamics near the onset of Covid 19 pandemic, so that we can compare the changes in dynamics with those during GFC. Our findings may be valuable as a supplement to other related research on the dynamics of large-scale catastrophic changes and help in the planning of strategies to reduce the likelihood of their occurrences.\\

Our paper is organised as follows: In Section \ref{sec2}, we outline the details of the data sets collected for our study and their processing. Section \ref{sec3} presents the methods of construction of recurrence plots and networks and their  measures useful in the analysis of the data. The variations in the measures and their implications are presented in Section \ref{sec4} and \ref{sec5}. The summary of our results and interpretations are presented in Section \ref{sec6}.

\section{Data sets and pre-processing }\label{sec2}
\begin{figure}
\centering
\includegraphics[scale=0.5]{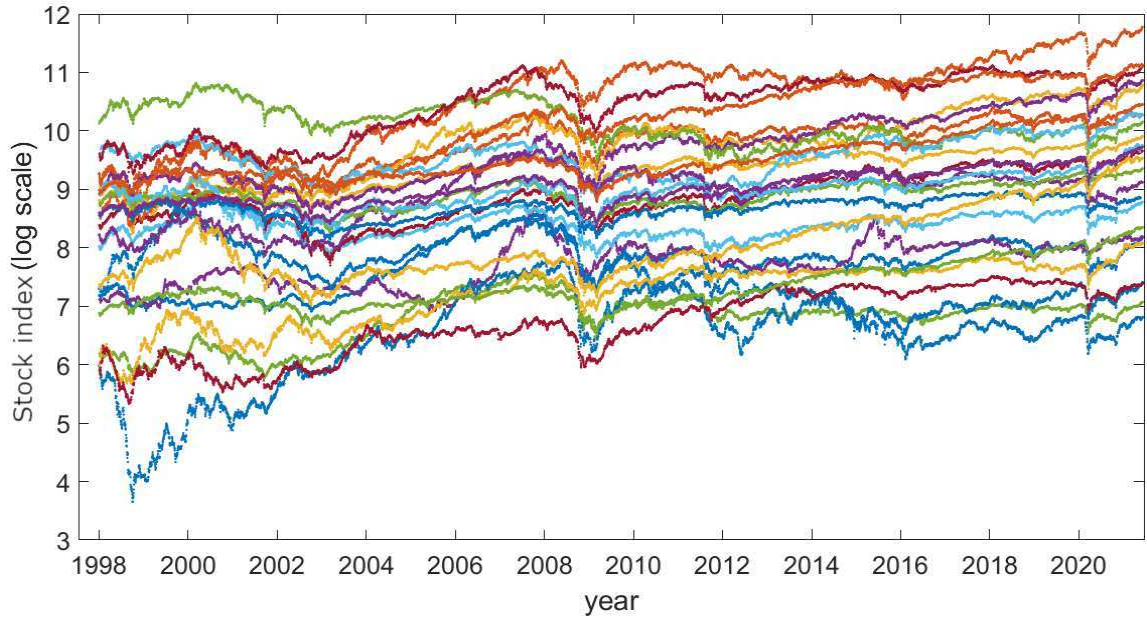}
\caption{Log plot of time series of stock market indices. Specific pronounced trends are clearly seen around 2008 and 2020 for all the markets.}
\label{fig:1}
\end{figure}

In our analysis we consider the daily closing prices of 26 international stock indices for a period spanning 2 January 1998 to 31 December 2021, from the publicly available source yahoo finance (https://in.finance.yahoo.com/), and the details are listed in Table \ref{tab1}. On days when the markets are closed, the index building process simply carry forward the index value from the previous business day. During crises, the variations in stock market indices are found to be significant, as is clear from Figure \ref{fig:1}, as pronounced trends in 2008 financial crisis and 2020 pandemic time. Hence we use the data of the closing prices rather than the more commonly studied returns. Since the listed stock indices includes major stock markets all over the world, we could also study the variations that happened in developed and emerging markets.

\begin{table}[H]
\begin{minipage}{174pt}
\caption{Market indices and symbols}\label{tab1}%
\resizebox{10cm}{!}{
\begin{tabular}{@{}lll@{}}
\toprule
Country 		&  Stock Index  			& Index Symbol\footnotemark[1] \\
\midrule
Austria			& Austrian Traded Index   		& ATX   \\
Brazil			& Bovespa Stock Index			& BOVESPA\footnotemark[2]  \\
Canada			& Toronto Composite Index		& SPTSX  \\
China			& Shanghai Composite Stock Exchange Index  & SSEC    \\
Czech Republic  & Prague Stock Exchange Index  	& SE PX   \\
France    		& CAC 40 Index   				& CAC   \\
Germany  		& DAX Index   					& DAX    \\
Greece    		& Athens Stock Exchange General Index & ASE    \\
Hong Kong   	& Hang Seng Index   			& HSI   \\
Hungary    		& Budapest Stock Index   		& BUX    \\
India   		& $S\& P $ BSE SENSEX Index   		& SENSEX   \\
Italy    		& FTSE MIB Index   				& FTSEMIB  \\
Japan   		& Nikkei 225 Index   			& NKY   \\
Poland    		& Warsaw Stock Exchange Index	& WIG   \\
Russia    		& MICEX Index   				& INDEXCF   \\
South Africa    & FTSE/JSE Africa All Share Index & JALSH   \\
South Korea		& Kospi Composite Index   		&  KOSPI  \\
Spain    		& IBEX 35 Index   				& IBEX   \\
Switzerland   	& Swiss Market Index   			&  SMI  \\
Taiwan    		& Taiwan Capitalization Weighted Stock index & TAIEX    \\
Thailand    	& SET Index   					& SET   \\
UK    			& FTSE 100   					&  UKX  \\
USA    			& Dow Jones Industrial Average  &  DJIA  \\
USA   			& NASDAQ Composite Index   		&  NASDAQ  \\
USA    			& NYSE Composite Index   		&  NYSE  \\
USA    			& $S\& P 500$ Index   			&  SPX  \\

\bottomrule
\end{tabular}%
}
\footnotesize{$^a$Index symbol is based on Bloomberg Indices.}
\footnotesize{$^b$Currently known as B3}
\end{minipage}
\end{table}

\begin{figure}
\centering
\includegraphics[scale=1.5]{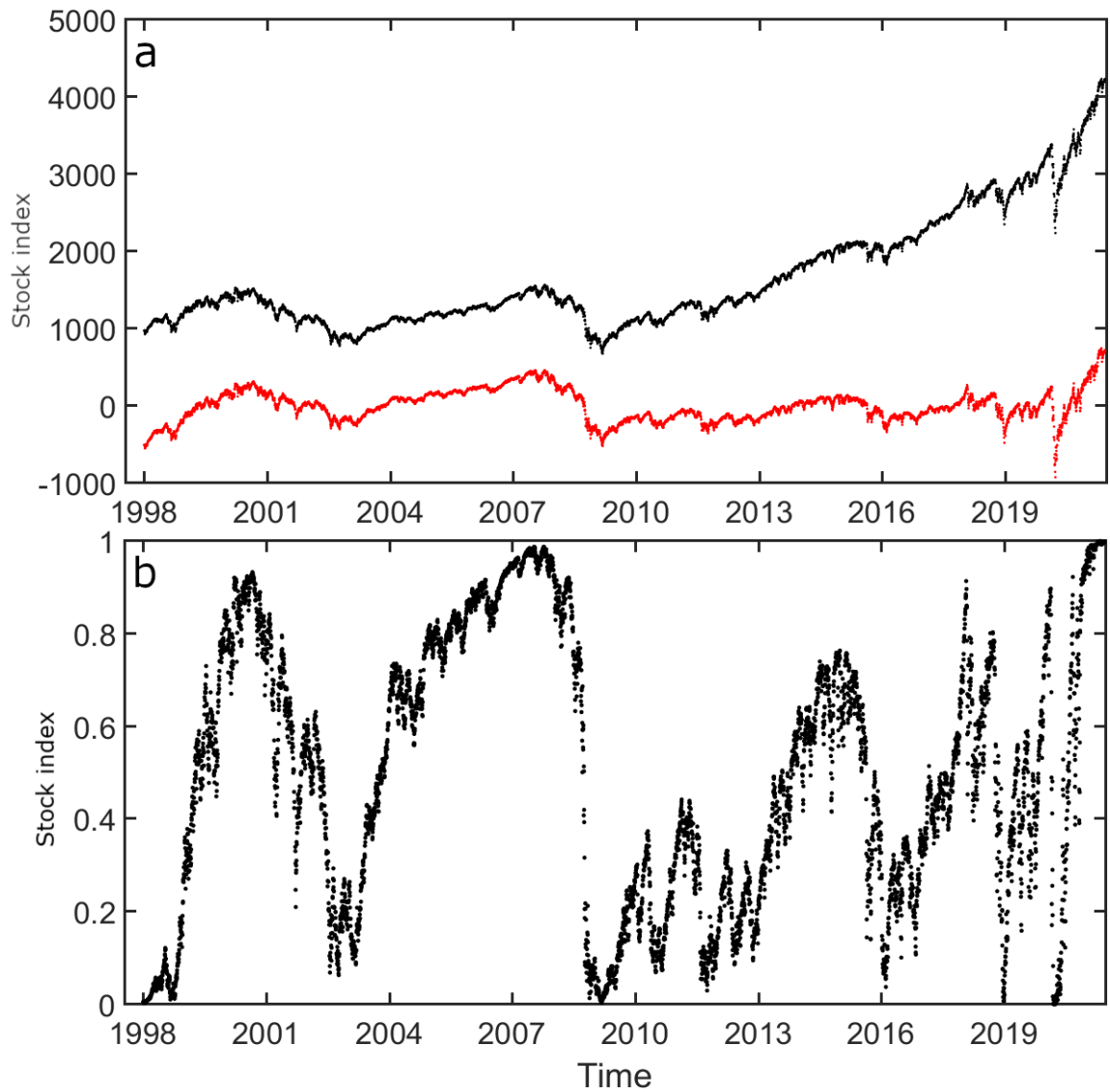}
\caption{(a)Time Series of a typical market, USA $S\&P 500$. The original time series is shown in black and the detrended time series in red.(b)Uniform deviate constructed from the time series of USA $S\&P 500$}
\label{fig:2}
\end{figure}

The first step in pre-processing of data is detrending using an appropriate polynomial fit to determine the global trend, which is then subtracted from the original signal to obtain the detrended signal \cite{Kotriwar2018HigherOS}. We confirm that detrending does not erase the finer structures in data, but only the long-term changes in the time series, as seen in Fig.\ref{fig:2} a for a typical data set.

We observe that the range of the data varies greatly from one market to the next, as can be observed from time series plots ( Fig.\ref{fig:1} ). So we apply the standard procedure of taking the uniform deviate of the detrended time series \cite{jacob2016uniform} to normalise all data values to the same range (0, 1). The data after taking uniform deviates is shown in Fig.\ref{fig:2} b for the data set of USA $S\&P 500$. 

\section{Methodology}\label{sec3}
\subsection{Recurrence Quantification}\label{subsec3}
Recurrence in dynamical systems is, defined as the occurrence of states that are arbitrarily close after a period of time, and is a fundamental property of bounded dynamical systems. 
The techniques using recurrence plots, collectively known as "recurrence quantification analysis" (RQA), have proven beneficial in characterising the behavior of underlying dynamics from data that are not sufficiently deterministic or stationary \cite{Marwan_2013}.
To reconstruct the state space trajectory, we embed the processed time series in an \textit{m}-dimensional space, using a suitable time delay $\tau$. Here both embedding parameters, the dimension m and the delay $\tau$, are to be chosen appropriately. Following the standard schemes, we take the time taken for the auto-correlation function to fall to 1/e as the time delay $\tau$ for that time series. The approach of false nearest neighbors(FNN) is used to estimate the suitable embedding dimension \cite{kennel1992determining}.\\
 
For the data sets used in the study, we find most of the markets have $\tau$ values around 200 days, while a few markets have very high $\tau$ value around 400. From calculations using FNN, we find the maximum value of embedding dimension, \textit{m}, among all the data sets is 4. For comparison across data sets, usually the maximum values is used for embedding all of them \cite{kantz2004nonlinear}. This is because the measures saturate at the required \textit{m} and so higher value of \textit{m} will not change the measures and their interpretations. Thus using \textit{m}=4, and the corresponding value of $\tau$, the phase space is reconstructed for each data.\\

From the reconstructed phase space, the recurrence plot (RP) is constructed  to visualise the recurrences in the dynamics. RP is a two-dimensional representation of its recurrences, where both axes are time, 
with dots depicting such recurrences within the neighbourhood $\epsilon$ of a state at time \textit{i} after a different time \textit{j}. It is represented as a Recurrence matrix, R defined, as
\begin{equation}
R_{i,j} =  \Theta(\epsilon -  \Vert x_i - x_j \Vert ) ,\; x \in \mathbb{R},\; i, j = 1...N
\end{equation}

where N is the number of considered states \textit{$x_i$}; $\epsilon$ is a threshold distance, $ \Vert . \Vert $ is a norm and $\Theta(.)$ is a Heaviside function. 

The value of $\epsilon$ is also fixed the same for all the data sets, as is usually done, for comparison across them \cite{10.3389/fnins.2021.787068,schinkel2008selection}. Here we take it to be 25\% of the range of values in each data. Since our data are all normalised in (0,1), $\epsilon$ is taken to be 0.25 for all the data sets.\\
We use two measures from RP, Determinism (DET) and Laminarity (LAM), for recurrence quantification analysis in the present study. Of these, determinism, defined as the fraction of recurrence points that form diagonal lines, is computed using, 
\begin{equation}
DET = \frac{\sum_{l=d_{min}}^{N}lH_D(l)}{\sum_{i,j=1}^N R_{i,j}}
\end{equation}
where $H_D(l)$ are the histogram of the lengths of the diagonal structures in the RP.\\
Laminarity (LAM) gives the percentage of recurrent points in vertical structures given by,
\begin{equation}
LAM = \frac{\sum_{l=v_{min}}^{N}lH_V(l)}{\sum_{i,j=1}^N R_{i,j}}
\end{equation} \\
where $H_V(l)$ are the histogram of the lengths of vertical structures in the RP.  

The results of our calculations show that all of the markets have very high DET and LAM values of more than $0.9$. The computed values of DET and LAM values in general, agree with already reported values \cite{bastos2011recurrence} and therefore are not reproduced here. We further try to characterise the dynamics using recurrence network measures.

\subsection{Recurrence Networks and Measures}\label{subsec4}
\begin{figure}
\centering
\includegraphics[scale=0.8]{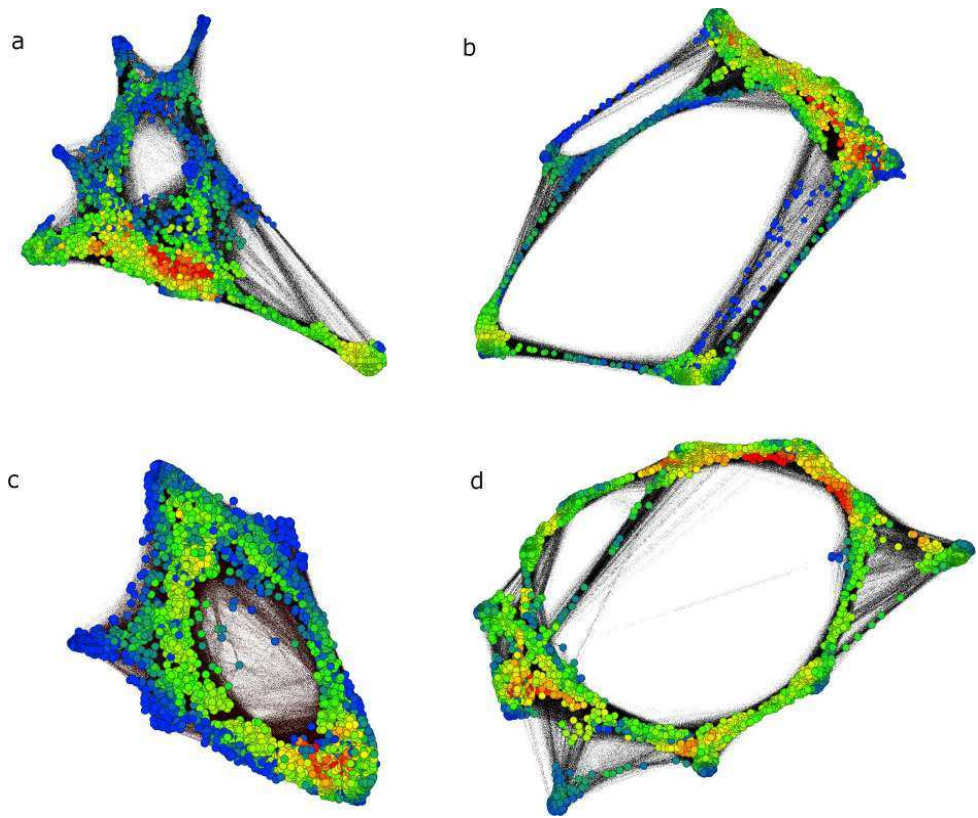}
\caption{Recurrence networks for the data sets of (a)USA $S\& P 500$, (b)France CAC40,(c)India BSE, (d)Japan Nikkei. The networks are plotted from adjacency matrix using GEPHI software (\href{https://gephi.org/}{https://gephi.org/}). The colour codes used relates to the degree of the nodes from blue to red, red colour showing highly connected nodes etc. }
\label{fig:3}
\end{figure}

From the recurrence matrix, the recurrence network(RN) is constructed by defining the adjacency matrix \( \textit{A} \) as
\begin{equation}
\textit{A}_{ij} = \textit{R}_{ij} - \delta_{ij}
\end{equation}

where $\delta_{ij}$ is the Kronecker delta \cite{marwan2009complex}.  Thus each point in the reconstructed phase space is considered as a node in RN and two nodes are connected by an edge if the distance between them in the embedded space is $\leq \epsilon$. The adjacency matrix A of the complex recurrence network is obtained by removing the self-loop (diagonal elements) from the matrix R. It is obvious that the matrix A is a binary, symmetric matrix implying that the resulting complex network is unweighted and undirected. The recurrence networks thus obtained  for four typical markets are shown in Fig. \ref{fig:3}. 

The standard complex network measures that can be used to understand the nature of the underlying dynamics from RN are Clustering Coefficient (CC) and Characteristic Path Length(CPL).

The CC quantifies the extent to which nodes in the network cluster together. For computing this, first the local clustering for a node $ \nu $ is computed as,
\begin{equation}
C_{\nu} = \frac{\sum_{ij}A_{\nu i}A_{ij}A_{j \nu}}{k_{\nu}(k_{\nu}-1)}
\end{equation}
Here $ A_{\nu i} $ are the elements of adjacency matrix. The average clustering coefficient CC for the network of size N is given by,
\begin{equation}
CC = \frac{1}{N} \sum_{\nu} C_{\nu}
\end{equation}
We also calculate Characteristic Path Length which is given by,
\begin{equation}
CPL = \frac{1}{N} \sum_{i}^{N} (\frac{1}{(N-1)} \sum_{i\neq j=1}^{N-1}l_{ij}^s)
\end{equation}
where $l_{ij}^s$ is the shortest distance between nodes \textit{i} and \textit{j}.

\begin{figure}
\centering
\includegraphics[scale=1.5]{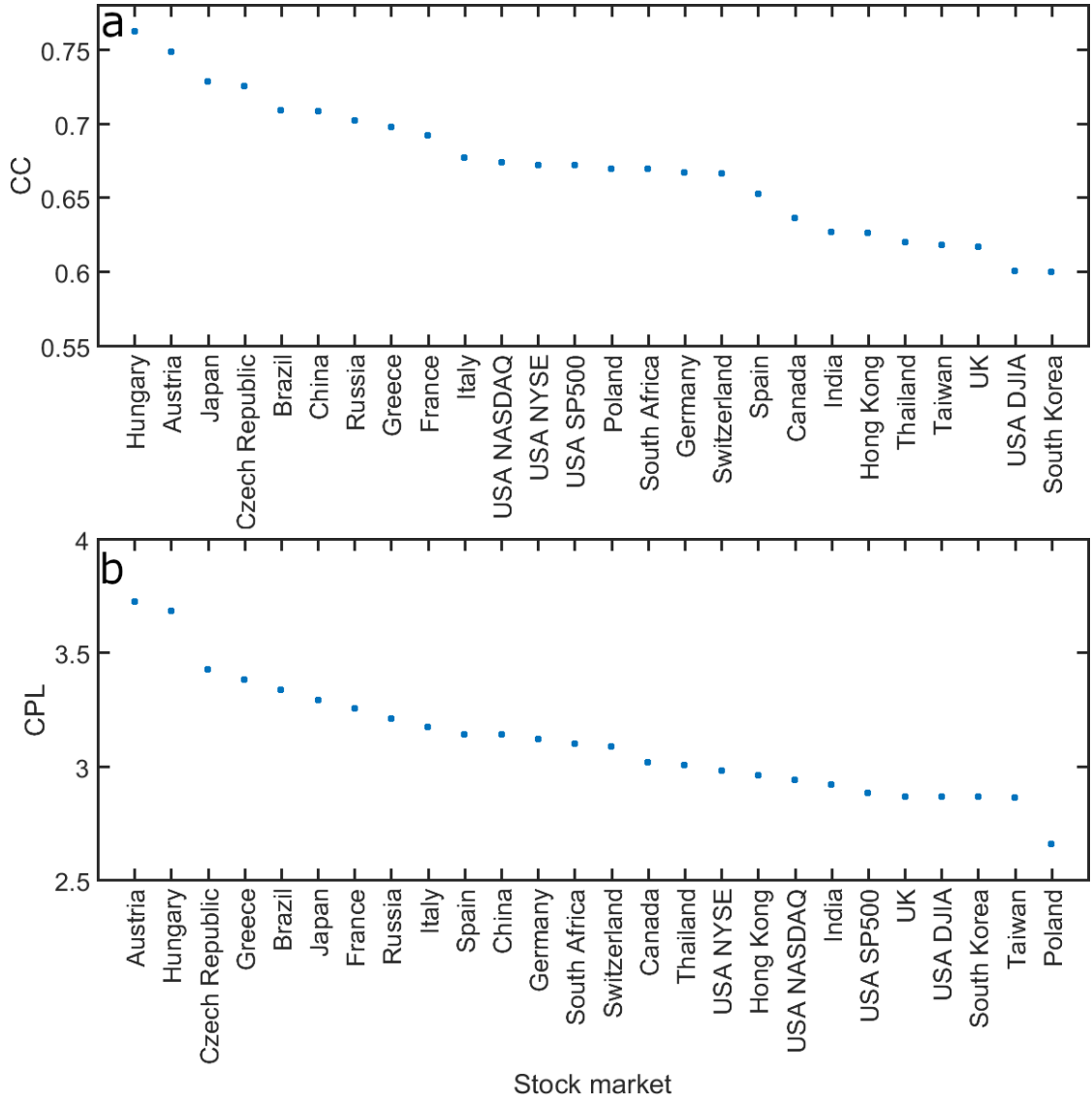}
\caption{(a)Clustering coefficient (CC) and (b)Characteristic path length (CPL) for recurrence networks for data of different stock markets}
\label{fig:4}
\end{figure}

The computed values of CC and CPL for different stock markets are shown in Figure \ref{fig:4}a and \ref{fig:4}b respectively, arranged in decreasing order of their values. In general these values indicate that all markets have underlying nonlinear dynamics. The almost flat regions in them indicate the groups among markets having the same values in CPL or CC. The stock markets of Austria ATX and Hungary BUX have high values of CC and CPL. Most of the developed markets has CPL values in the range [2.8, 3], while emerging markets like Greece ASE, Czech Republic SE PX, Brazil BOVESPA lie in the relatively higher CPL range. The Asian markets, such as India BSE, Hong Kong HSI, and Taiwan TAIEX, have CC values in the range of [0.6, 0.65] and developed markets like the United States $S\& P 500$, Germany DAX, and Switzerland SMI, fall within [0.65, 0.7] in CC values.

\section{Transitions in Stock market Dynamics}\label{sec4}

In addition to fluctuations in values, the stock markets are reported to undergo major changes or regime shifts over time. They are in general classified as bubbles and crisis but can be further classified into different phases. 
Asset prices that significantly vary from their core values are typically referred to as economic bubbles. An example of reported bubble is the dot-com bubble in US around 2002. And the crash of US housing market is usually referred to as the burst of the US Housing Bubble. Thus, Bartram and Bodnar \cite{bartram2009no} investigated 2008 financial crisis, and mentioned that at the beginning of Oct 2007 world equity markets were all time high, whereas by the end of Feb 2009 equity markets dropped off more than $56 \%$. They proposed Jan 1, 2007 to Sep 12, 2008 as pre-crisis period and Oct 28, 2008 to Feb 2009 as the post-crisis period. Mishkin \cite{mishkin2011over} divided the financial crisis into two distinct phases. The first phase from Aug 2007 to Aug 2008 called the US Subprime mortgage crisis. The second phase in mid Sep 2008 is called the Global Financial Crisis. \cite{pisani2010banking,dooley2009transmission,claessens2010cross}, from 2007 to 2009.\\

To understand such transitions or regime shifts as dynamical transitions in the underlying dynamics of stock markets, we use  the sliding window analysis over the data for the period 1998-2017. The window considered in our study is 1500 time steps long, and we slide it by 100 time steps. Then we embed the data within each window, and get the corresponding recurrence plots and networks. We compute recurrence network measures CC and CPL as well as recurrence plot measures DET and LAM. The values obtained for each window is assigned to the center point in that window to get the temporal evolution of the recurrence measures. Their variations over time can indicate the changes in the underlying dynamics that can be captured from the time series of stock market data. 

\subsection{Trends in CPL and CC}\label{subsec7}

\begin{figure}
\centering
\includegraphics[scale=1.5]{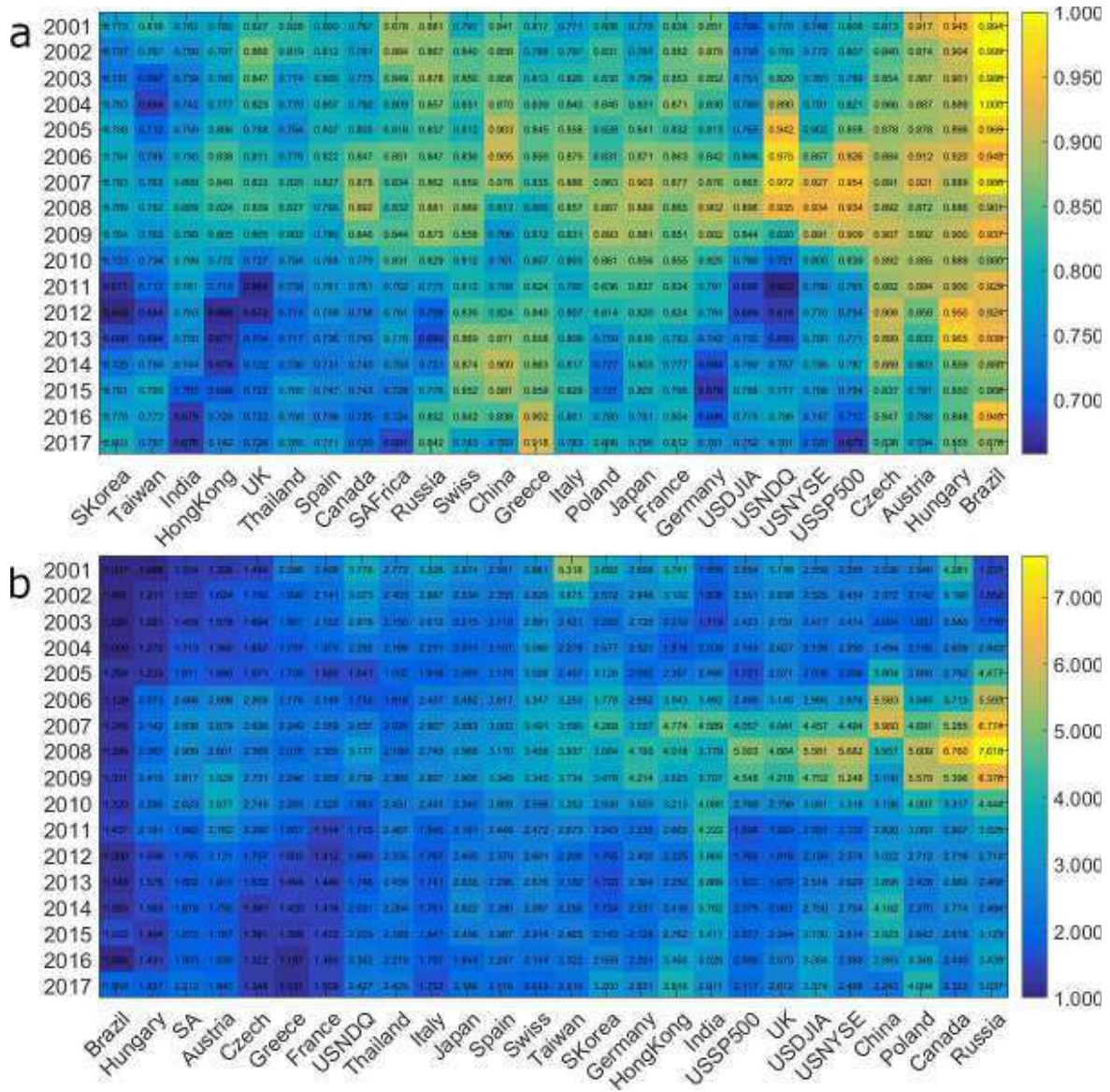}
\caption{Heat map of (a)CC values and (b)CPL values for data sets of stock indices. The markets that are more affected during the global financial crisis are shown to the right.}
\label{fig:5}
\end{figure}

The changes in CPL and CC values over time for all the stock market indices is represented as a heat map in Fig:\ref{fig:5} with markets arranged in the increasing order of their changes. Thus the markets on the left show moderate changes in the values of CC and CPL over time while those on the right are largely affected. 

\begin{figure}
\centering
\includegraphics[scale=0.8]{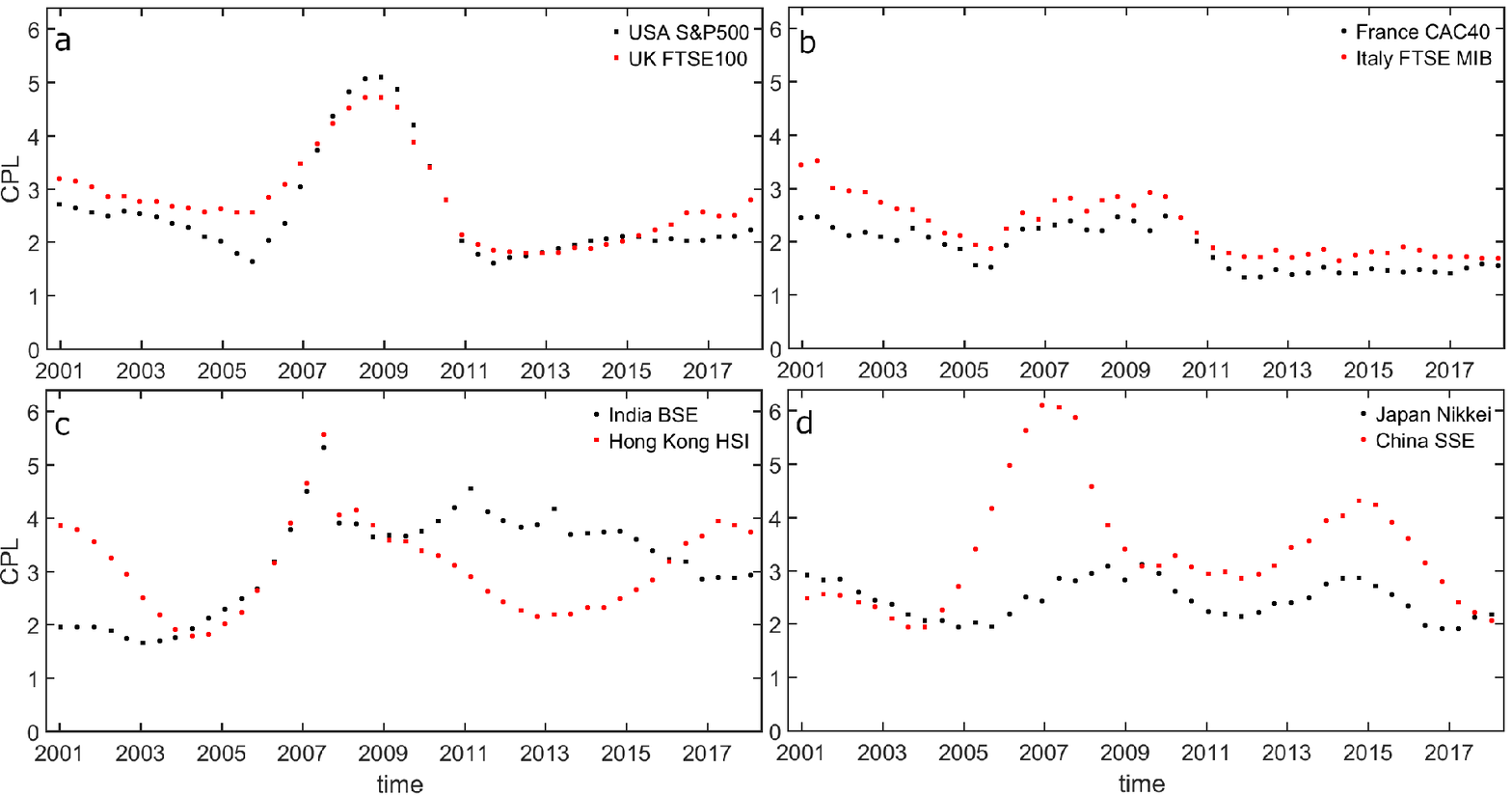}
\caption{Characteristic Path lengths from sliding window analysis for recurrence networks constructed from data sets of (a)USA $S\& P 500$, (b)France CAC40,(c)India BSE, (d)Japan Nikkei. CPL is high prior to 2008 and decreases as the crisis approaches.}
\label{fig:6}
\end{figure}

We find the trends in the variations in CC values are not that pronounced  but are better revealed through CPL values. For further analysis we  consider a few typical markets and show the variations in their CPL values grouped as per the nature of their variations(Fig. \ref{fig:6}).\\

We note that USA $S\& P500$ and UK FTSE100 show an increasing trend in CPL values starting from 2006 , reaching a maximum around 2008, followed by a decrease till 2011. Compared to these markets the increase in CPL started earlier, around 2005,  in the Asian markets, India BSE and Hong Kong HSI,  start decreasing by end of 2007 .  The changes in China SSE is much steeper soon after 2004, reaching maximum around 2007 and decreasing till end of 2009. But changes in European markets France CAC40 and Italy FTSE MIB, and Japan Nikkei are much less pronounced with a small rise during 2007-08 and decrease during 2009-2010.
Moreover China SSE and Japan Nikkei rise and fall in CPL values during 2013-2014, which can be correlated to the currency crash reported around the same time.
 
We infer that decrease in CPL values can be associated with stochasticity entering the dynamics, where people started to buy and sell the stocks randomly. Hence increasing trend in CPL values could be considered as an indication of stock markets approaching a crisis.\\

\subsection{Trends in DET and LAM}\label{subsec8}

From the same sliding window analysis, we compute the two measures, DET and LAM, from their recurrence plots (RP). The decrease in DET and LAM appear to be ahead of crash, which is a notable feature of their evolution. During the 2008 crash events, significant declines in the levels of DET and LAM were reported earlier also \cite{bastos2011recurrence}. We discuss this further in the next section.

\section{Transitions in Dynamics during GFC and pandemic times }\label{sec5}
In this section we compare the dynamics of stock markets during the 2008 GFC and the pandemic event. We check for critical transitions and possible early warning signals during these two events. As reported the reason for the GFC is because, between 2001 and 2003, the US Federal Reserve decreased its discount rate a total of 27 times \cite{lin2008impact}. Low interest rates fuelled significant credit growth, aided by massive trade surpluses that China and other countries exploited to buy US Treasury Bonds. House price increases accompanied by significant credit expansion, particularly through mortgage lending. Sub-prime mortgage lending to homeowners without the necessary means to repay loans skyrocketed in the United States. By the summer of 2007, rising mortgage defaults and foreclosures in the United States had signalled that the sub-prime market was in crisis. \\

The COVID-19's impact on stock market has been labelled as a "black swan event" by some economists, referring to the occurrence of a highly unexpected event with catastrophic consequences. The rapid spread of the novel corona virus COVID-19 since January 2020 and the subsequent lockdowns have caused instability in stock markets \cite{albuquerque2020resiliency}. The occurrences of volatility clustering were observed in Asian markets during the pandemic \cite{mishra2020corona,narayan2020japanese}. The increase in COVID-19 confirmed cases, changes in oil prices, and unexpected lockdowns were found to be significant causes of fears and uncertainties of the pandemic to cause unexpected decline in global stock markets price.\\

\begin{figure}[H]
\centering
\includegraphics[scale=1.5]{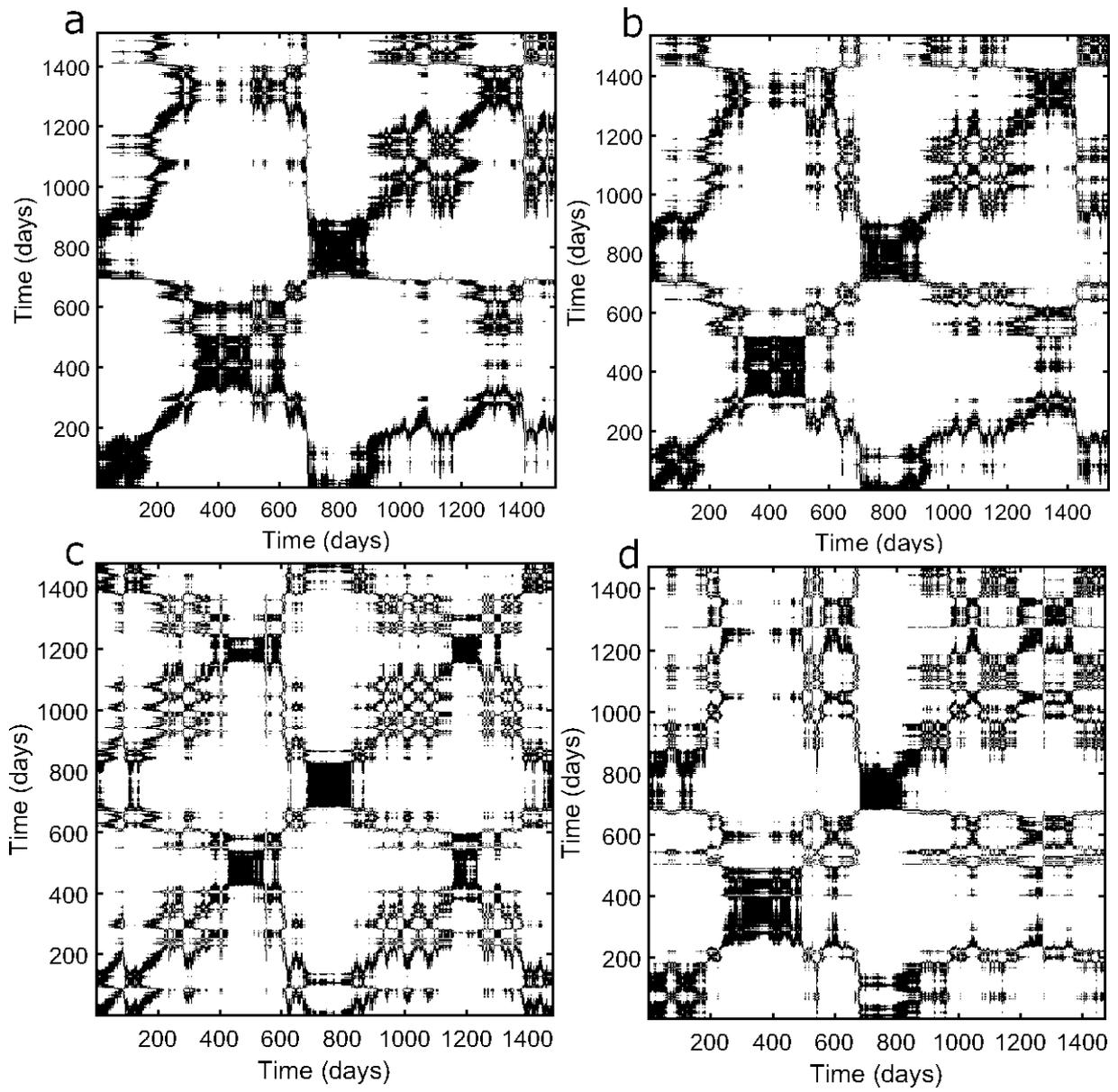}
\caption{Recurrence plots during GFC for data sets of (a)USA $S\& P 500$, (b)France CAC40,(c)India BSE, (d)Japan Nikkei.}
\label{fig:7}
\end{figure}

\begin{figure}[H]
\centering
\includegraphics[scale=1.5]{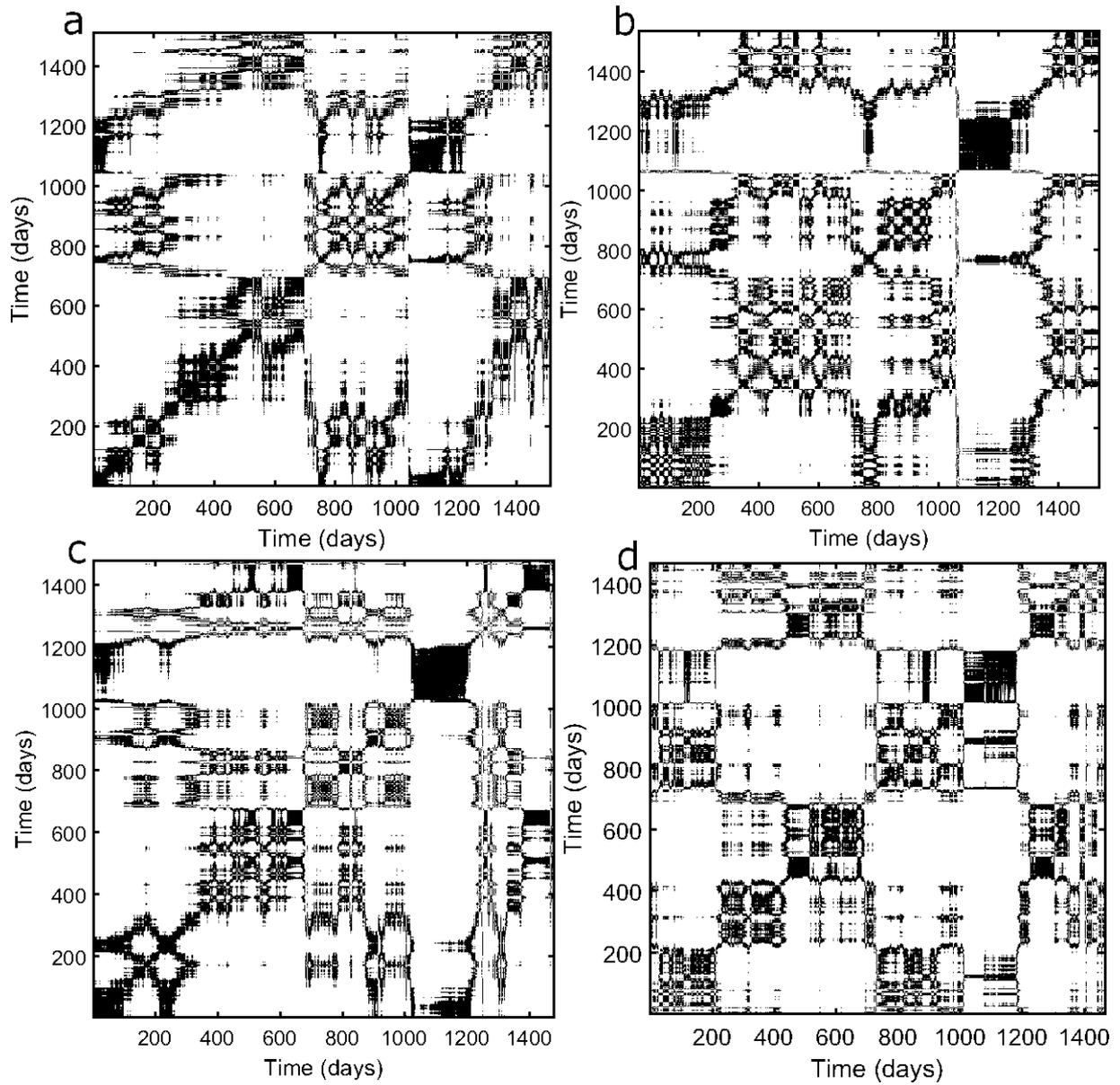}
\caption{Recurrence plots during the pandemic for data sets of (a)USA $S\& P 500$, (b)France CAC40,(c)India BSE, (d)Japan Nikkei.}
\label{fig:8}
\end{figure}

\begin{figure}[H]
\centering
\includegraphics[scale=0.8]{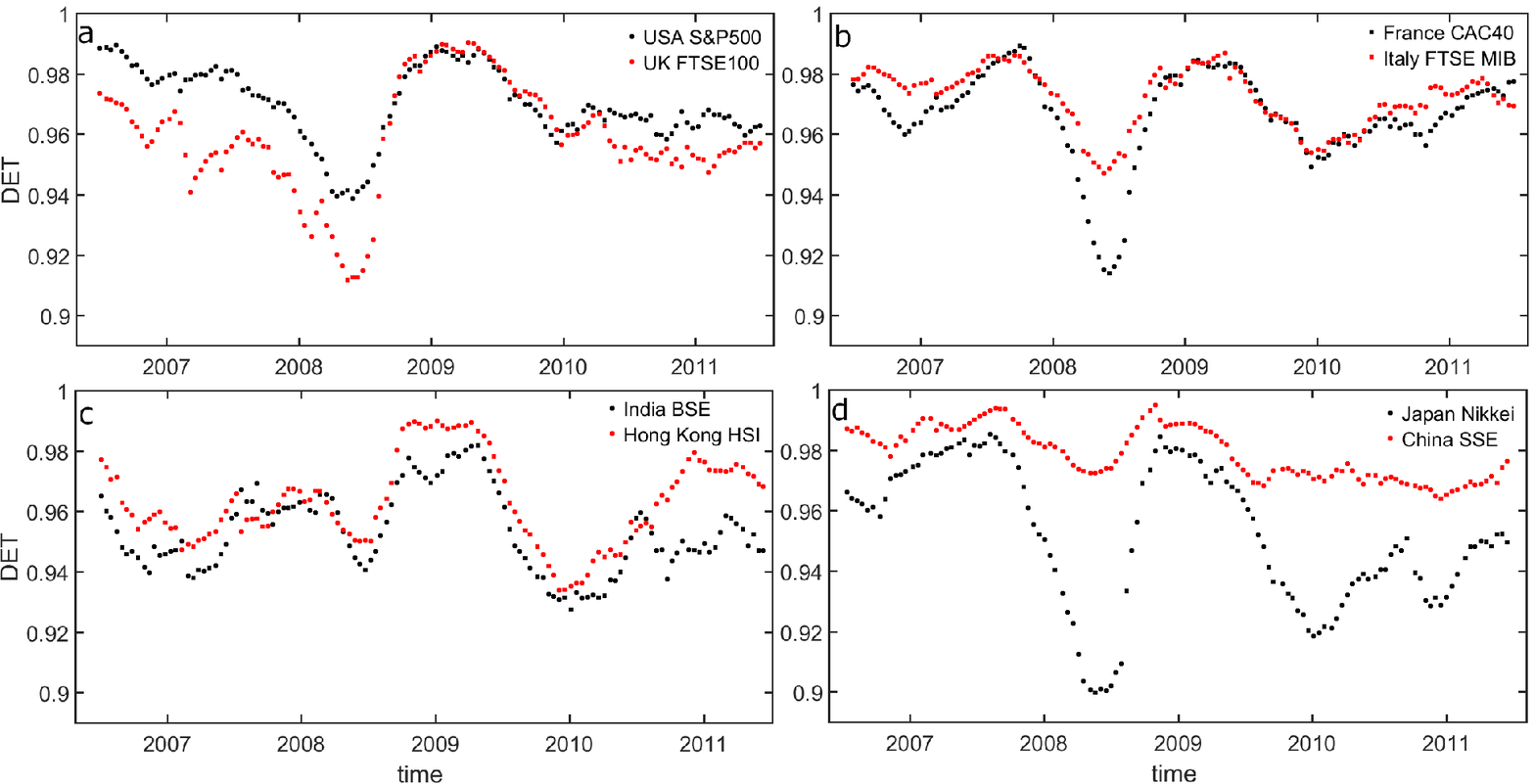}
\caption{Variation of DET over time during GFC for data sets of (a)USA $S\& P 500$ and, UK FTSE100, (b)France CAC40 and Italy FTSE MIB, (c)India BSE and Hong Kong HSI, (d)Japan Nikkei and China SSE.}
\label{fig:9}
\end{figure}

\begin{figure}[H]
\centering
\includegraphics[scale=0.8]{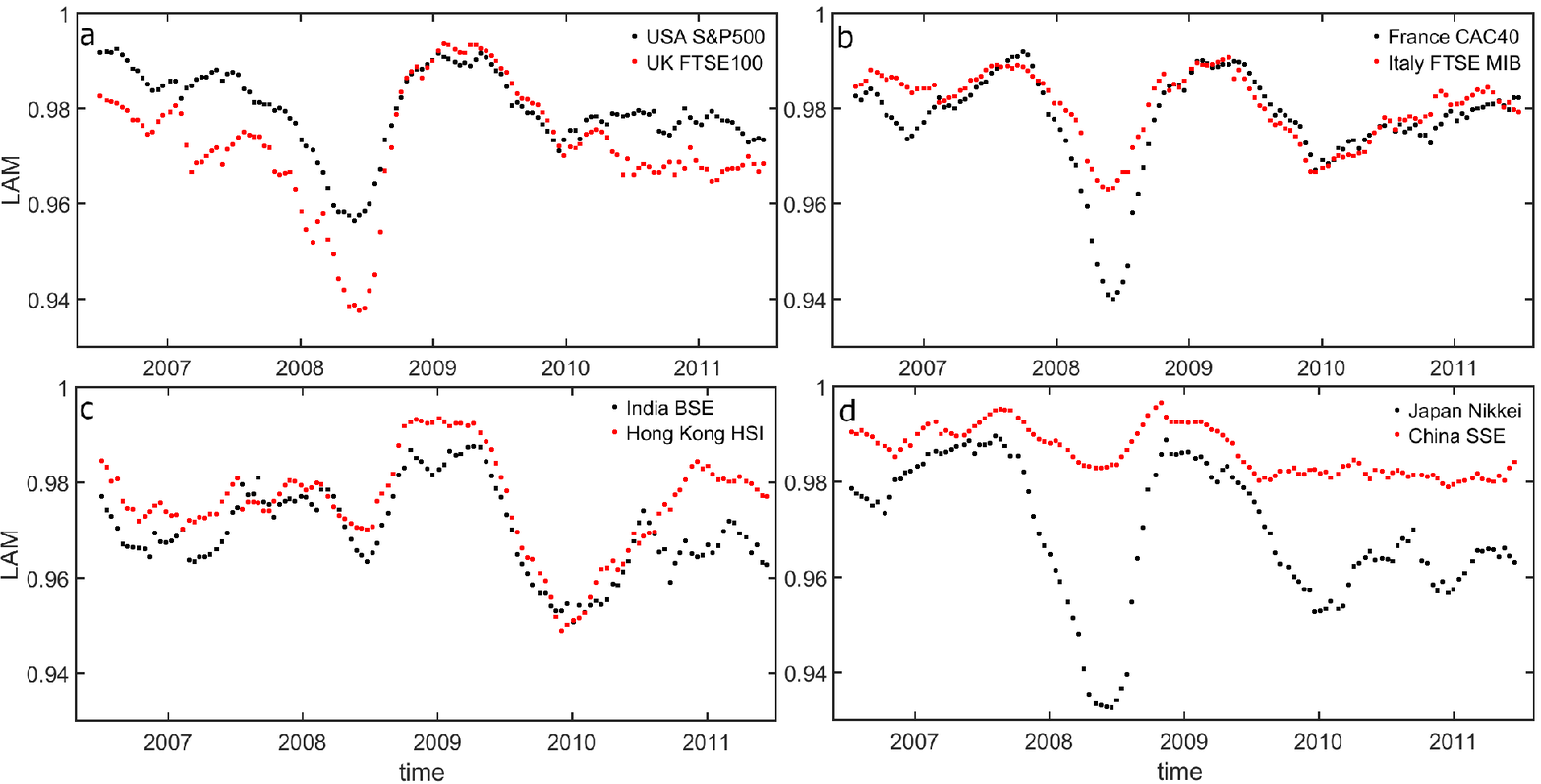}
\caption{Variation of LAM over time during GFC for data sets of (a)USA $S\& P 500$ and, UK FTSE100, (b)France CAC40 and Italy FTSE MIB, (c)India BSE and Hong Kong HSI, (d)Japan Nikkei and China SSE.}
\label{fig:10}
\end{figure}

\begin{figure}[H]
\centering
\includegraphics[scale=0.8]{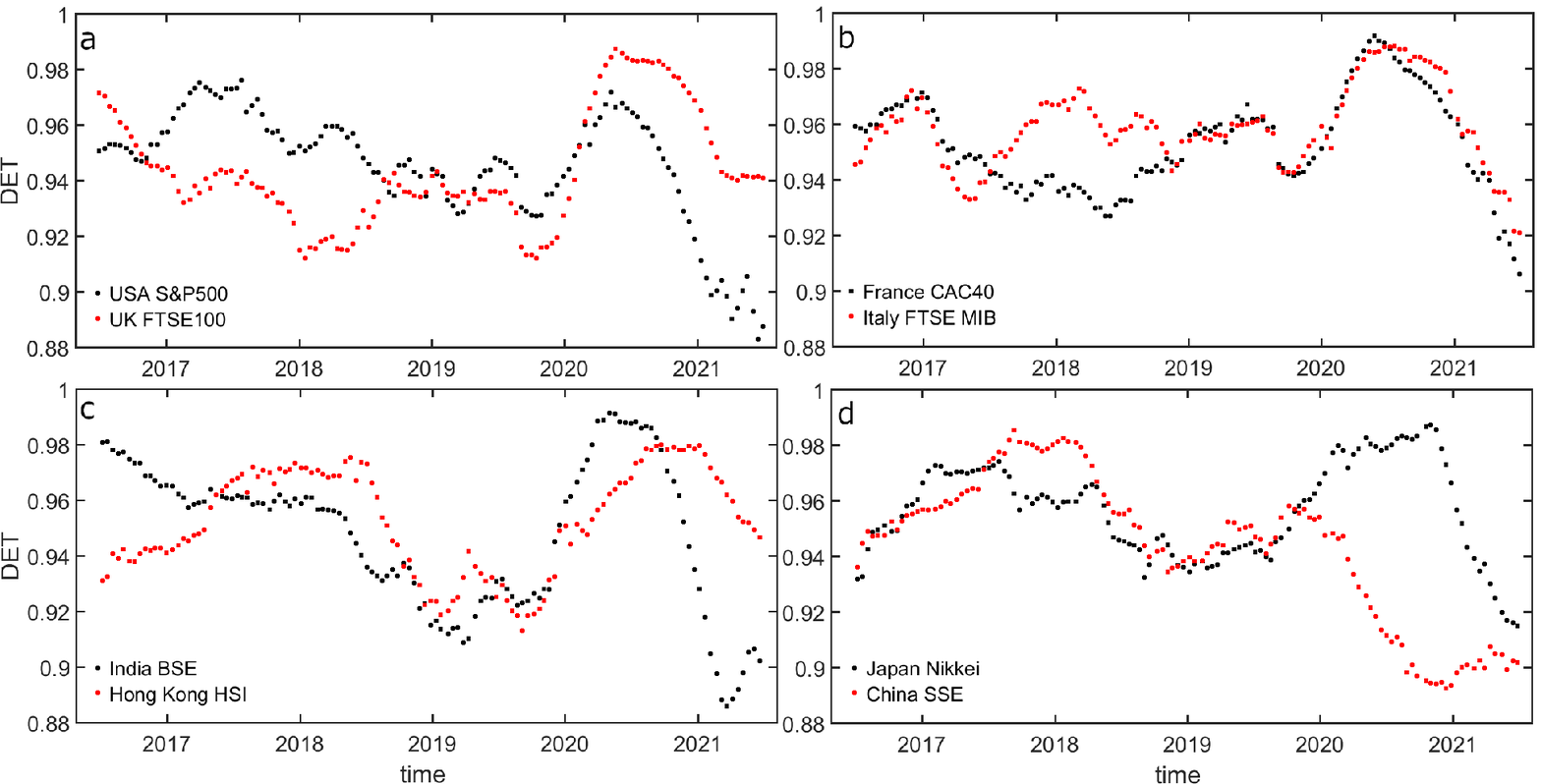}
\caption{Variation of DET over time during pandemic for data sets of (a)USA $S\& P 500$ and, UK FTSE100, (b)France CAC40 and Italy FTSE MIB, (c)India BSE and Hong Kong HSI, (d)Japan Nikkei and China SSE.}
\label{fig:11}
\end{figure}

\begin{figure}[H]
\centering
\includegraphics[scale=0.8]{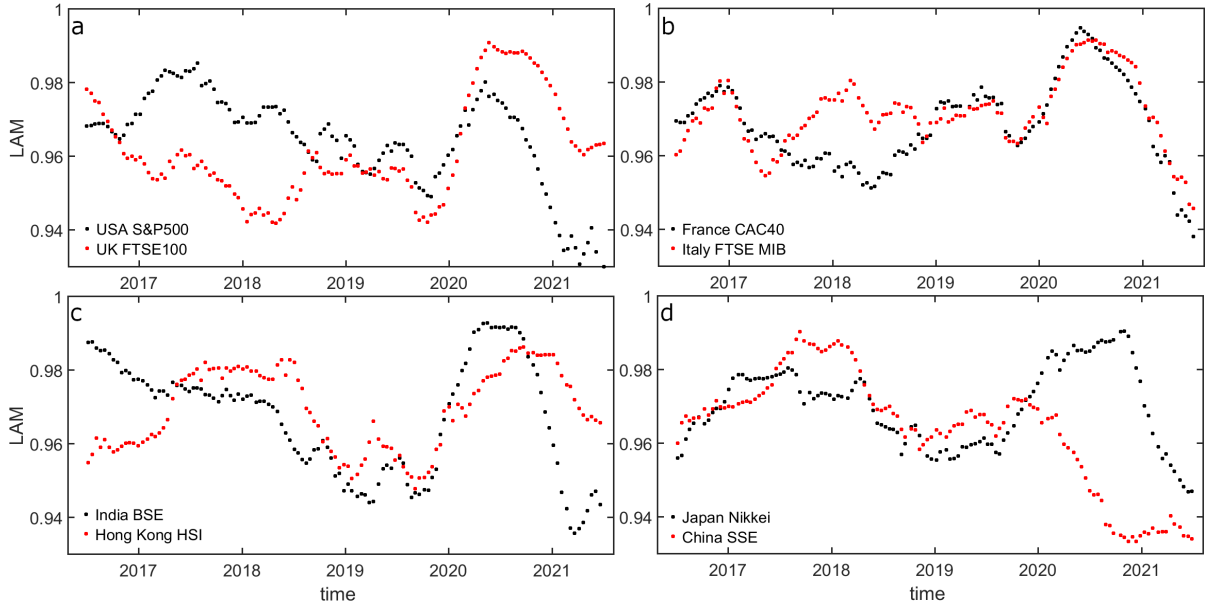}
\caption{Variation of LAM over time during pandemic for data sets of (a)USA $S\& P 500$ and, UK FTSE100, (b)France CAC40 and Italy FTSE MIB, (c)India BSE and Hong Kong HSI, (d)Japan Nikkei and China SSE.}
\label{fig:12}
\end{figure}

We conduct sliding window analysis for the data during the period from 2006 to 2011 for the GFC and 2016 to 2021 for pandemic. Since the size of data during the recent pandemic related time series is less, we have to use sliding window of size 250 and slide by 10 points. For comparison, we repeat computations for the GFC also with the same window size and slides. Also the recurrence plot is arrived at without embedding as was done in similar recent studies \cite{ngamga2016evaluation,goswami2013global,george2020early,martin2018improving}. The RPs thus obtained for typical markets are shown in Fig. \ref{fig:7}  during GFC and those during pandemic in Fig  \ref{fig:8}.
We present a comparison of values of recurrence measures that can indicate the changes in dynamics during the GFC and recent pandemic related event.
For this, we compute the recurrence measures, DET and LAM, using sliding window analysis during these two periods and are presented in figures:\ref{fig:9}, \ref{fig:10}, \ref{fig:11} and \ref{fig:12}.\\

We find both the measures show almost the same trends during GFC. For most of the markets, the values of DET and LAM start decreasing from the start of 2008 reaching a minimum by start of 2009. Thereafter they increase to peak by middle of 2009, then decreases to stabilise. The changes in India BSE and Hong Kong HSI are slightly different with a delayed steeper minimum in mid 2010 before increasing again. For Japan Nikkei also, the values continue to decrease till mid 2010 but decrease during pre-crisis period is more.
The trends in the values indicate that the dynamics becomes less deterministic during the time approaching the critical event and so DET and LAM can be used as precursors of critical transitions. The decrease in DET and LAM can be linked to increasing stochasticity in the dynamics. These indications are seen in changes in CPL values also during the same period.However during the pandemic time, our study indicates increase in DET/LAM soon after start of 2020, and the values start decreasing at different times for the markets. The China SSE was the first to show decrease in DET/LAM by mid 2020 while for Japan Nikkei, the decrease started only by mid 2021. India starts decreasing trend by end of 2021, reaching a minimum and starts to increase after that while Hong Kong HSI starts to decrease only by end of 2021.  USA $S\& P 500$ continues decreasing while UK FTSE100 has stabilised. France CAC40 and Italy FTSE MIB is delayed in responses during the pandemic, increase starts later in 2020 and decrease by start of 2021. Comparing with the trends in measures around GFC, we can say the trends seen in pandemic time can be interpreted as a crisis that can happen in recent future.\\

\begin{figure}[H]
\centering
\includegraphics[scale=0.8]{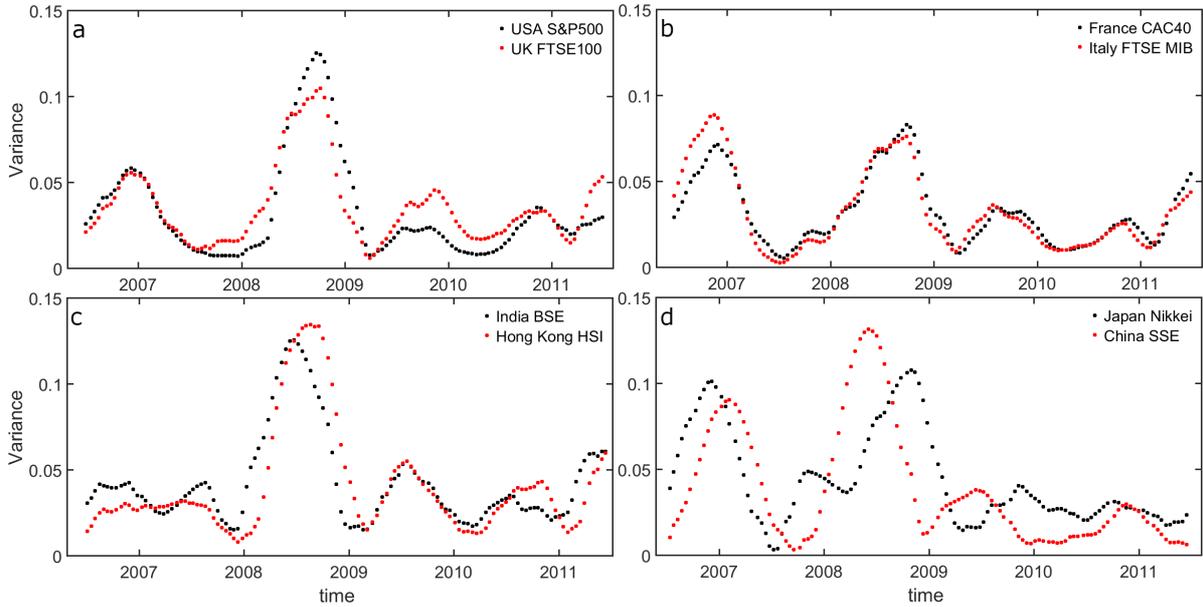}
\caption{Variance over time during GFC for data sets of (a)USA $S\& P 500$ and, UK FTSE100, (b)France CAC40 and Italy FTSE MIB, (c)India BSE and Hong Kong HSI, (d)Japan Nikkei and China SSE.}
\label{fig:13}
\end{figure}

We now discuss the standard statistical measures used as early warning signals in critical transitions in complex systems. As reported in \cite{guttal2016lack}, prior to the crash of 2008, detrended time series reveal that autocorrelation at lag-1, a key indicator of critical slowing down, showed either no or weak trends, as measured by the Kendall-$\tau$ rank correlation coefficient. The variance of the time series is another widely used
statistic for indications of critical transitions and the rise in variance near thresholds is observed in a variety of social and ecological systems \cite{brock2006tipping}, \cite{carpenter2006rising}. So we compute variance using sliding window analysis for all markets to check indications of approaching transitions.\\
An increasing trend of variance for the US $S\& P500$, UK FTSE100 and German DAX markets prior to the crash of 2008 was reported by Guttal \cite{guttal2016lack} which is similar to the results obtained (Fig.\ref{fig:13}).
This shows shifts in trends at the same time intervals and increase in variance matches with decrease in DET/LAM and vice versa.
These results support our earlier inference that there are indications of increasing stochasticity creeping into the dynamics, before approaching GFC.\\

We confirm the trends observed further by applying modified Mann-Kendall test \cite{HAMED1998182}, keeping p value $ < 0.05$ for the 8 markets chosen for detailed analysis. During the pre-crisis period (June 2007 - Aug 2008) we find DET and LAM show significant decreasing trend in all markets, except India BSE and Hong Kong HSI for which the trends are less significant with the above criteria. During this period the increase in variance is significant for markets UK FTSE100, France CAC40, Italy FTSE MIB, India BSE, Japan Nikkei and China SSE, which further confirms our inferences. During the crisis, LAM and DET start increasing but post crisis (Sept 2008-Sept 2010), DET and LAM shown significant decreasing trends in all markets and variance shows a decreasing trend for most of the markets except  UK FTSE100, India BSE and Japan Nikkei.\\

\begin{figure}[H]
\centering
\includegraphics[scale=0.8]{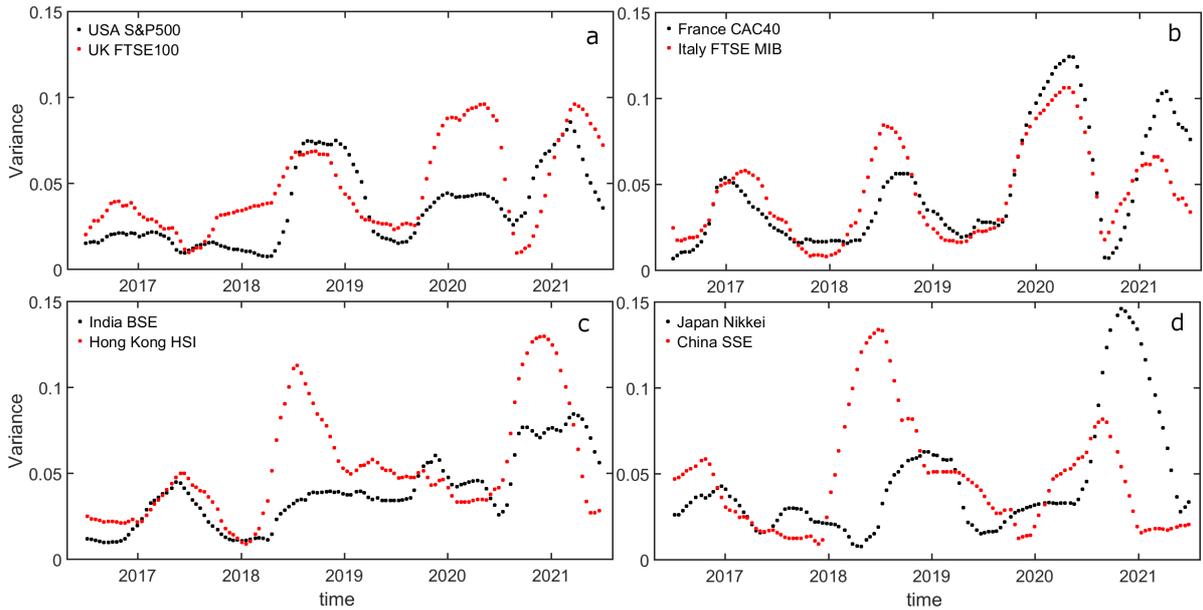}
\caption{Variance over time during pandemic for data sets of (a)USA $S\& P 500$ and, UK FTSE100, (b)France CAC40 and Italy FTSE MIB, (c)India BSE and Hong Kong HSI, (d)Japan Nikkei and China SSE. }
\label{fig:14}
\end{figure}

We analyze the pandemic time, (July 2020-July 2021) using the same test and the markets are found to show significant decreasing trends except China SSE, for which the decrease starts earlier. The trend in variance is not significant in this context except for UK FTSE100, that shows a slight increasing trend (Fig.\ref{fig:14}). Prior to this period, variance shows decreasing trend for Italy FTSE MIB and increasing trend for India BSE.

\section{Summary and Conclusion}\label{sec6}

In this study we present a detailed analysis of the dynamics of stock markets that can be inferred from their data. For this we reconstruct the dynamics using the data of stock market prices from 26 markets spread over the globe and compute their recurrence based measures. Our main motive is to understand the changes in their underlying dynamics from these measures focusing on the transitions during the 2008 Global Financial Crisis(GFC) and the pandemic related changes after 2020. From the recurrence networks constructed from the dynamics, we compute the clustering coefficient (CC) and characteristic Path length (CPL) for the whole data which indicate the nonlinear nature of their dynamics and identify markets that can be grouped with similar values of CPL and CC.\\

To understand the transitions in dynamics during regime shifts, we do a sliding window analysis covering the period 1998-2017 and obtain the heat maps for the values of CC and CPL. With these values, we arrange the markets in order of increasing variations in CC and CPL. Moreover, for typical markets we show in detail the changes in CPL during the GFC. It is clear that CPL values show an increasing trend before GFC and suddenly start decreasing during GFC. Once markets stabilised, they returned to pre-crisis values. The difference in the time of onset of crisis among markets also could be captured in the analysis.  The increase in CPL can be related to a dynamical state that is more stochastic or noisy in nature that happens before the crisis till 2008. Our results align well with the earlier reported study that CPL is high during noisy or turbulent regime and decreases as periodic regime is approached \cite{godavarthi2017recurrence}. Hence CPL can be used to indicate the proximity of impending transitions from data.\\

We extend our study to compare the nature of transitions around GFC and pandemic time using the recurrence plot based measures, DET and LAM. We could identify decreasing trends in the variations of DET and LAM before the crash reaching a minimum by end of 2008, followed by a sudden increase till 2009. This also indicates that stochastic nature creeps into the dynamics before the crisis. This is also  supported by the trends in the variance calculated from data. Thus our results on CPL values, DET and LAM measures confirm that the GFC is dominated by stochastic driven dynamics. The origin of these stochastic fluctuations is not very clear but can be due to random selling and buying driven by panic and other drastic measures.\\

The changes seen in DET and LAM before the pandemic prior to 2019 can be related to Brexit, trade war, crack down on black money by India etc. These are reported to have transient effects in markets during 2016-2019. During this period, the values of DET and LAM decrease and reach a minimum before 2019. They start increasing, showing a tendency to stabilise, when the markets were suddenly hit by the unexpected pandemic. We can see that DET and LAM start decreasing fast since decreasing from mid 2020 for most of the markets, while for China SSE, it starts earlier. During this time, we also find the trend in variance is not significant in general with variations specific to markets. This can be due to multiple random external influences that can vary for each markets. However results with data till 2021 indicate that they have not yet stabilized for most of the markets. Therefore this  could mean the approach of a crisis in the near future.\\

Some of the interesting results of the study are the differences in the nature of transitions and responses of market during GFC and pandemic time. Thus the markets that go together during GFC are responding differently in pandemic time. This is to be expected since the underlying causes and mechanisms are different during these two events.

The role of recurrence measures as early warning signals for critical transitions has been studied earlier in different contexts like thermoacoustic instability and dimming of Betelgeuse etc.. Our study establishes their application to understand transitions and regime shifts in financial markets.  The recurrence quantification is applicable for short and non-stationary data sets, and their measures are computationally faster. This makes them especially useful and efficient in many practical applications to understand the onset of major regime shifts from the recurrence patterns in the data. Our study brings out the importance of stochastically driven transitions in stock market dynamics with either internal or strong external random influences. As such this will have relevance in other complex dynamical systems that undergo transitions but may not show critical slowing down.


\end{document}